# On Critical Infrastructures, Their Security and Resilience - Trends and Vision

Craig Rieger, Milos Manic

Idaho National Laboratory[1], Virginia Commonwealth University[2],

Craig.Rieger@inl.gov[1], misko@ieee.org[2]

*Abstract*—This short paper is presented in observance and promotion of November, the National Month of Critical Infrastructure Security and Resilience (CISR), established by the United States Department of Homeland Security in 2013 [1]. The CISR term focuses on essential assets (critical infrastructures) and two ultimate goals of making them secure and resilient. These assets and goals were put together in 2013 in the now well-known Presidential Policy Directive on CISR (PPD-21) [2]. This paper presents easy-to-ready material laying down the building blocks of CISR—what it means to you as a regular citizen, professional, or government worker. This paper presents concepts behind security and resilience pertinent to various types of activities—from every day to field-specific activities. This paper also presents basic elements to the field: (1) high-level introduction to the organizational units dealing with CISR in the United States; (2) explanation of basic terms and a list of further reading material; and (3) several discussion topics on the vision and future of CISR in critical infrastructure cyber-physical systems.

## I. Critical infrastructures, their security, and resilience

Let us first break down the term "critical infrastructure security and resilience (CISR)." Critical infrastructure (CI) includes the assets or infrastructures, relevant to everyday life—from the power in our homes, the means of transportation we take to and from work, to the food and water we depend on daily. Security and resilience refers to the goals of making CIs secure and able to recover upon failures of various natures.

**What are critical infrastructures (CIs)?** CIs are our nation's backbone. They refer to both physical and cyber systems vital to our nation's physical or economical security, health, and safety. CIs carry different meaning to different people. The United States (U.S.) Department of Homeland Security [3] cites the now very well-known PPD-21, the Presidential Policy Directive on CISR, [2] from February 2013. PPD-21 defines CIs and what security and resilience of CIs means. PPD-21 identifies 16 CI sectors (e.g., food and agriculture, healthcare, dams, energy, financial services, government facilities, and communications). These infrastructures entail assets, systems, and networks (physical or virtual), vital to the United States. PPD-21 also states that federal government should engage with international partners to strengthen the security and resilience of CIs outside of the United States on which the nation depends.

Security and resilience of CIs advances a national policy to strengthen and maintain secure, functioning, and resilient CIs.

**What is security and resilience?** The Cyber Infrastructure Security Agency (CISA) [4] defines security as reducing the risk of natural or manmade disasters to CIs. This includes both physical and cyber threats. Physical measures, such as fencing, guarding, and cyber measures such as intrusion detection systems and antivirus software measures are used to provide protections to cyber-physical threats.

Resilience measures could be anything from backup power generators, business continuity plans, to software tools for anomaly detection and recovery [17].

In recognition of ever-increasing challenges in securing CIs, just this month (November 13, 2018), the U.S. House of Representatives voted unanimously to create CISA within the Department of Homeland Security. The CISA Act reorganizes the current National Protection and Programs Directorate into a new agency and prioritizes its mission as the federal leader for cyber and physical infrastructure security. CISA appropriately entails two divisions, the Cybersecurity Division and the Infrastructure Security Division [5].

**Resilience, the new term?**

**Interdependency:** As noted at The First NSF Workshop on Resilience for Critical Infrastructures, at National Science Foundation, organized by the authors of this paper, one of the key aspects of resilience in infrastructures is interdependency [6]. The level of sophistication of current infrastructures implies high levels of interconnectedness and interdependency (e.g., hospitals depend on electrical energy, which relies on transmission lines, part of a power grid, drawing power from power plants). In the case of Hurricane Katrina, the supply of crude oil was interrupted because of a loss of electric power for three major transmission pipelines, leading to a loss of 1.4 million barrels of oil per day supplies (accounting for 90% of the production in the Gulf of Mexico).

Examples of interdependency can be very simple. For example, a backup generator being on top of a building may be positioned too high to bring gas necessary for its operations. Similarly, a backup generator placed in a basement may be too susceptible to flood. For these very reasons, directive PPD-21 specified that U.S. efforts should address security and resilience of infrastructures in an integrated and holistic manner to reflect the infrastructure's interconnectedness and interdependency [2].

**Interdisciplinary approach:** As interconnectedness and interdependency were noted at Resilience Week, one of the key aspects of resilience in infrastructures is interdependency [7], which requires an interdisciplinary approach to address. Moving from a multidisciplinary approach, where different disciplines unite to perform their complementary role, an interdisciplinary approach evolves the interpersonal dynamic to a collective understanding and language of resilience that efficiently advances to optimal solutions.

**Resilience, definitions and meaning:** The vocabulary and formality of the term resilience is constantly evolving [8]. From resilience in the workplace (capacity to respond to pressures of daily life), to resilience of physical systems (ability to spring back, recover, even adapt to stress and threats). In epidemiology, the term dates back to the early 70s [9]. In realm of control systems, a description of a resilient control system was established as *the system that maintains state awareness and an accepted level of operational normalcy in response to disturbances, including threats of an unexpected and malicious nature* [10] (Figure 1). Understandably, the definition resilience is adapted to the areas where resilience is being investigated. In cyber-physical systems, it entails physical, cyber, and cognitive domain [11], [12]. This implies human systems, cybersecurity, complex control, and/or networked systems, as well as communications among systems.

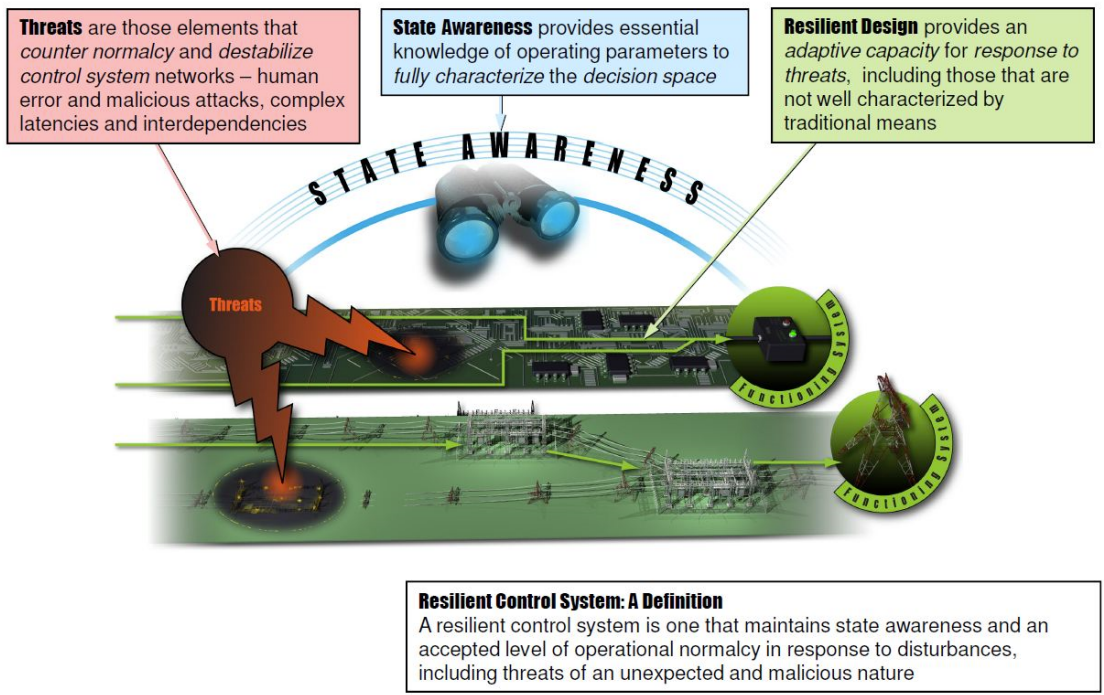

Figure 1. Resilient control system.

**Ability to recover, the bio inspiration:** In contrast to *security* (methods and processes of *protecting* data or physical resources), *resilience* brings the fundamental aspect of *ability to recover*. A frequently used quote referring to cyber-attacks states that *it is not a question if, but rather when it will happen*. Here, the resiliency refers to the reality—whether your organization or control system in question has the ability to withstand and/or quickly recover from threats, benign or malicious, to a state that enables continued functionality. The ability to recover naturally has been "mimicked" using artificial intelligence (AI) or machine learning techniques. Hence, AI approaches known as fuzzy logic, neural networks, and genetic algorithms have been applied to network security or energy systems [13], [14]. One of the noted strengths of deploying AI in resiliency is the fact that most AI approaches are data driven—in modern infrastructures, heavily interconnected and interdependent, massive data sets are being collected and stored. Hence, researchers have successfully applied data-driven techniques to health monitoring of cyber-physical systems CPS [15], or *autonomic intelligent cyber sensors* in industrial control networks [16], [17], [18].

## II. TRENDS AND VISION

**Resilience metrics:** In order to evaluate, strengthen, or predict resiliency of a certain system, resilience metrics need to be defined. This could be the amount of energy per unit volume required to stress a material to *yield stress limit*. Recent work on *resilience metrics* integrates the cognitive, cyber-physical aspects that should be considered when defining solutions for resilience [19], [20], [21] (Figure 2). Physics-based approaches provide the best representation of performance when considering resilient design tradeoffs and correlating uncertainty. However, physics-based approaches are not always readily definable for every application, and data-driven approaches can be applied, especially in cases of uncertainty propagation where physics-based approaches are not defined or where historic knowledge can inform the decision process [31].

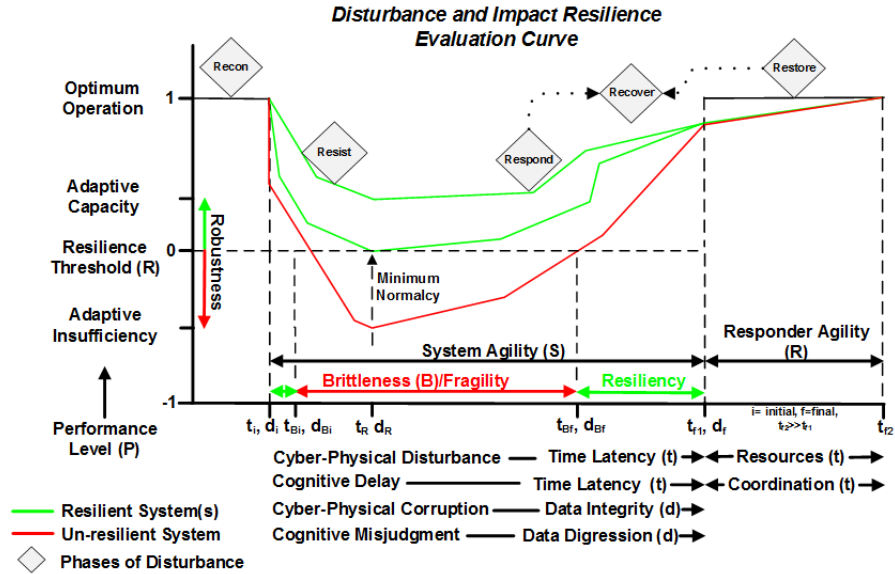

Figure 2. Resilience base metrics.

**Deep learning (DL) and massive data sets:** In modern scenarios of massive data sets being collected, deep learning (DL) or deep neural networks emerge as a tool of "general intelligence." Such are applied to modern eco-systems such as cyber aware, DL powered, and human interacting intelligent buildings of the future [12], [16], [23]. In modern AI systems, one of the key goals is *generalization of DL*—the ability of a DL model to adapt to the behavior of CPS to previously unseen scenarios in cybersecurity and resilience [24].

**Trust, explainability, adversarial learning:** One of hot topics in DL and resilience recently has been increasing trust into AI-based cyber and resilience systems [25]. While initial ideas on explainable AI (XAI) are credited to DARPA [26], just months ago researchers successfully showed how deep neural-network-based intrusion detection systems can be increasingly helpful in improving user trust [27]. Adversarial learning is yet another exciting emerging field—exploiting how a DL system can be "fooled" to wrong conclusions—knowing which can strengthen the system against incorrect intrusion detection decisions, and hence increase trust and explainability of such systems [28] (Figure 3). As the future seems to be bringing the age of machines attacking machines, researchers have very recently shown how a classifier can be fooled to miscategorize a panda as a gibbon or a turtle as a rifle by changing a single pixel only [29].

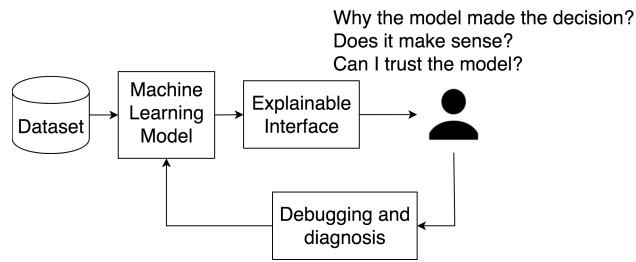

Figure 3. Explainable models and trustworthy AI.

**Multi-agent approaches to complex dynamic architectures:** The ability to address the complexity of control system dynamics and the underlying need for resilience has suggested advancement of multi-agent approaches. The computer science and AI community have developed the initial basis for agent-based architectures, but more recently the control system community has started considering the unique considerations of dynamic systems in this light. While some approaches consider the approach in terms of time-based dynamics, the reality of control systems includes human involvement in the management and coordination of the system [22], [30] (Figure 4). This hybrid control scenario provides a rich research opportunity, advancing both cognitive processing that would provide methods to capture optimized human teaming behaviors, and AI through algorithmic correlation of these behaviors for use in a multi-agent hierarchy.

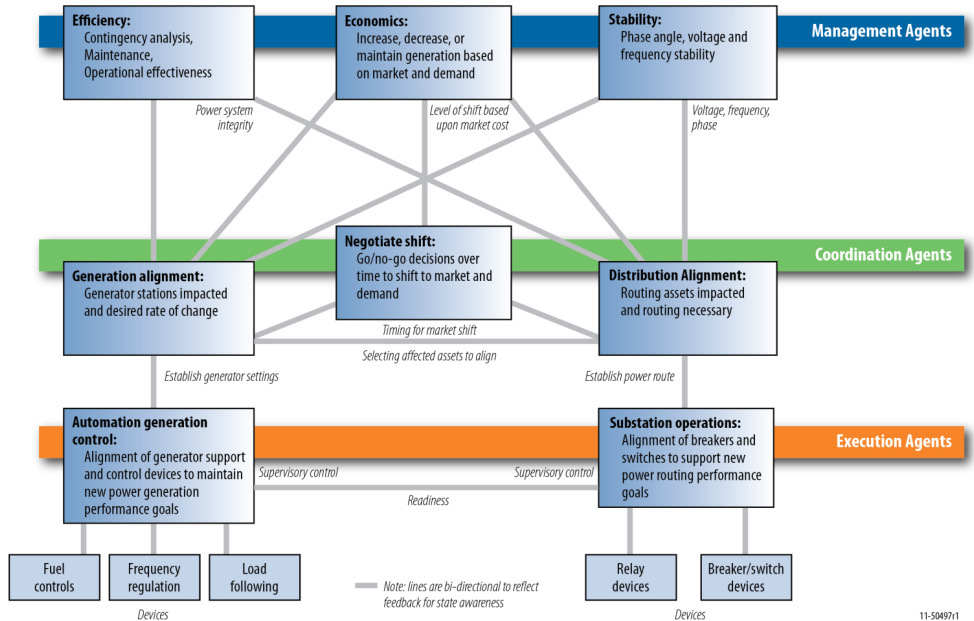

Figure 4. Notional Multi-agent Architecture Decomposition for Power Grid

## III. Summary

This paper does not attempt to provide a comprehensive survey of issues regarding security and resilience in CIs. Rather, it merely tries to open up discussion on some of the emerging topics.

Resilience continues to evolve since early 70s, but with increasing researcher and practitioner interest in the new formalisms that represent the resilience benefit. It is obvious that interdependency and interconnectedness is the key in evaluating and strengthening of security and resilience of modern control systems, which entail, but are not limited to, physical, cyber, and cognitive aspects. This diverse nature of threats, therefore, requires an evolution from multidisciplinary to interdisciplinary teams to develop the language and formalism. Here, resilience metrics remains one of the highly researched topics.

With recent technology advancements and prevalence in cyber-physical and Internet of Things (IoT) devices, massive data sets become the norm where traditional approaches to security and resilience seem to be increasingly shifting towards modern AI techniques. As with obvious benefits of modern AI systems, questions of machine versus machine wars arise. Researchers continue to recognize and tackle challenges of AI in the future—in particular the problems of understanding and formalizing trust and explainablity with AI.

To engender resilience into distributed control systems of the future, decomposition of complex control designs suggest a decomposition of the system into optimally-stabilizable entities with limited, definable interactions such as an agent-based approach. In addition, the age-old hybrid control challenge finds potential in considering AI approaches to address the event-based human decisions and resulting agents that drive control system performance basis.